# Electronic film with embedded micro-mirrors for solar energy concentrator systems

**Mario Rabinowitz*[a]** and **Mark Davidson[b]**

April 16, 2004

[a]Armor Research, Redwood City, CA 94062-3922  USA    Mario715@earthlink.net
[b]Spectel Research Corp., Palo Alto, CA 94303  USA    mdavid@spectelresearch.com

**Abstract**
      A novel electronic film solar energy concentrator with embedded micro-mirrors that track the sun is described.  The potential viability of this new concept is presented.  Due to miniaturization, the amount of material needed for the optical system is minimal.  Because it is light-weight and flexible, it can easily be attached to the land or existing structures.  This presents an economic advantage over conventional concentrators which require the construction of a separate structure to support them, and motors to orient them to intercept and properly reflect sunlight.  Such separate structures must be able to survive gusts, windstorms, earthquakes, etc.  This concentrator utilizes the ground or existing edifices which are already capable of withstanding such vicissitudes of nature.

## 1.  Introduction

Solar energy has proven to be as elusive as it is alluring because the power we receive from the sun isn't concentrated, being only about 1.4 kW/m$^2$ at its highest above the atmosphere.  On earth it is 1 kW/m$^2$ , at best, on a sunny day.  Despite problems of pollution and relentless resource attrition, fossil fuel furnaces provide high power density giving them an attractive appeal if one isn't concerned about long term consequences.  In contrast, the promise of pollution-free renewable solar energy can be achieved by utilizing modern micro-fabrication and micro-electronics to focus sunlight from a large area onto a small area by means of a relatively inexpensive, thin, flexible, electronic film attached to the ground or existing structure.

At this time all forms of Solar Energy represent only a small percentage of the power generated in the U.S. Presently it costs 2 to 5 times more to generate electricity using solar than from fossil fuels with current production costs of electricity, of about 2 to 4 ¢/kWh (van der Zwaan and Rabl). Consumer costs are much higher. Conventional concentrator-PV systems may at best cost about 6 to 10 ¢/kWh (Swanson, 2000). The electronic film concentrator can have much lower costs.

**2. Concentrating sunlight**
Since the solar energy flux is low, concentrating it seems like a natural thing to do. A large number of concentrator technologies have already been proposed (Swanson, 2000). The analysis of solar power systems are complicated by a large number of variables (Duffie and Beckman, 1974; Rabl, 1985). The problem with conventional concentrator systems is that they are not much better than unconcentrated systems for the following reasons:

1. Structural requirements for worst-case weather conditions.
2. Cost of high quality optical surfaces.
3. Cost of machinery (motors and gears) for orienting the collectors.
4. Cost of computers for tracking and selectively orienting collectors.
5. Cost of installation, maintenance, and cleaning.

**3. Electronic film solar tracking solution**
We propose a solar concentrator that has very low mass because it is so thin, and yet has the toughness and durability of mylar sheet. Thus it can easily be attached to existing structures such as buildings or the ground. It utilizes the type of technology that has been developed for electronic paper such as the gyricon concept (Sheridon and Berkovitz, 1977) and (Sheridon et al, 1999). This new kind of electronic film has small rotatable mirrors embedded in it which can be rotated and controlled by electronic means (Davidson and Rabinowitz, 2003). Solar concentrators must withstand harsh winds, gusts, windstorms, earthquakes, etc. Meeting these requirements adds significantly to the cost of traditional concentrators, but not ours. In the concept proposed here, the structural strength is already built into an existing edifice, and so the cost of the concentrator is greatly reduced. It can be used to concentrate sunlight incident on the film at any angle onto a small area where a solar photovoltaic surface or other energy conversion device can utilize the increased flux of solar light in order to produce electricity or heat.

There are many variations of this basic idea. In one variation the individual mini-reflectors consist of spheres, one hemisphere of which is transparent, with a mirror in the equatorial plane (cf. Fig. 1). This sphere would be small, ~ 200 microns (0.2 mm) in diameter, and would be suspended in a lubricating liquid that would allow it to rotate freely. The liquid would be index of refraction matched to the clear hemisphere, and it should have the same density as the sphere to minimize buoyant forces. The spheres should be roughly balanced to minimize gross gravitational orientation . The spheres can be manufactured so as to have either an electric dipole moment (electret), a magnetic dipole moment (magnet), or both, and these electromagnetic dipole moments can be used to control the orientation in tracking the sun.

In the technology developed by Xerox in the 1970's referred to as Gyricon (Sheridon and Berkovitz, 1977) and (Sheridon et al, 1999) , the electric dipole moment of a spherical ball made from two dielectrically different hemispheres originates from the shear-interaction of the sphere with the fluid that it is immersed in. Shear results as motion is induced by electrostatic grid interaction with an embedded electret shown in the left side of Fig. 1. The spheres have a surface charge, with equal and opposite charge in the adjacent liquid, called the electrical double layer. The "zeta potential," is the net surface and volume charge that lies within the shear slipping surface resulting from the motion of the ball through the liquid (Sheridon and Berkovitz, 1977) and (Hunter, 1981). The zeta potential is an electrical potential that exists across the interface of all solids and liquids. It is also known as the electrokinetic potential. Figure 1 shows how the zeta potential can lead to a net electric dipole moment. Charges are shown only for the transparent material, although similar charge layers (approximately of equal and opposite sign) are formed on the gray back hemisphere's surface as well.

Electrets with volume electric dipoles are commercially available and are commonly used in sensitive microphones. The old problem of maintaining charge separation seems to be well in hand. Electrets are now almost as stable as magnets (Sessler, 1998). The strength of the dipole moment does not need to be the same for each ball as it relates to the angular acceleration of a ball rather than its orientation.

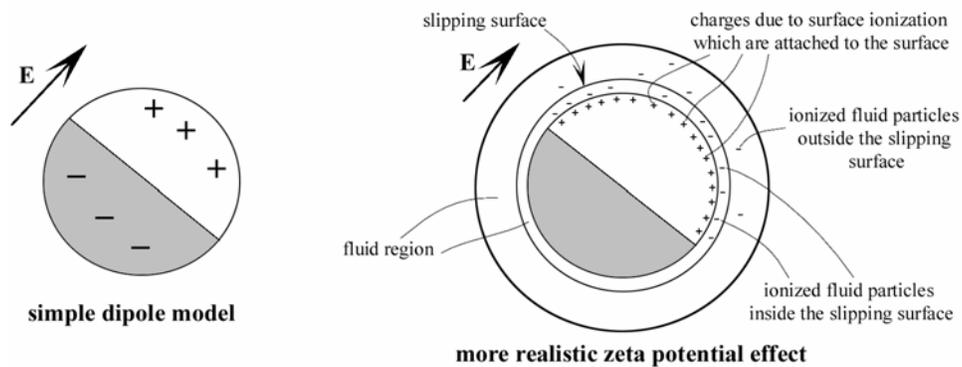

**Figure 1.** Gyricon Twisting-Ball Model Showing a Simple Embedded Dipole and an Induced Zeta Potential Dipole.

Spheres such as that in Fig. 1 would fill a monolayer (Fig. 2) or more. The electric dipole moment of each sphere can be oriented by electric fields in the gyricon technology (Sheridon and

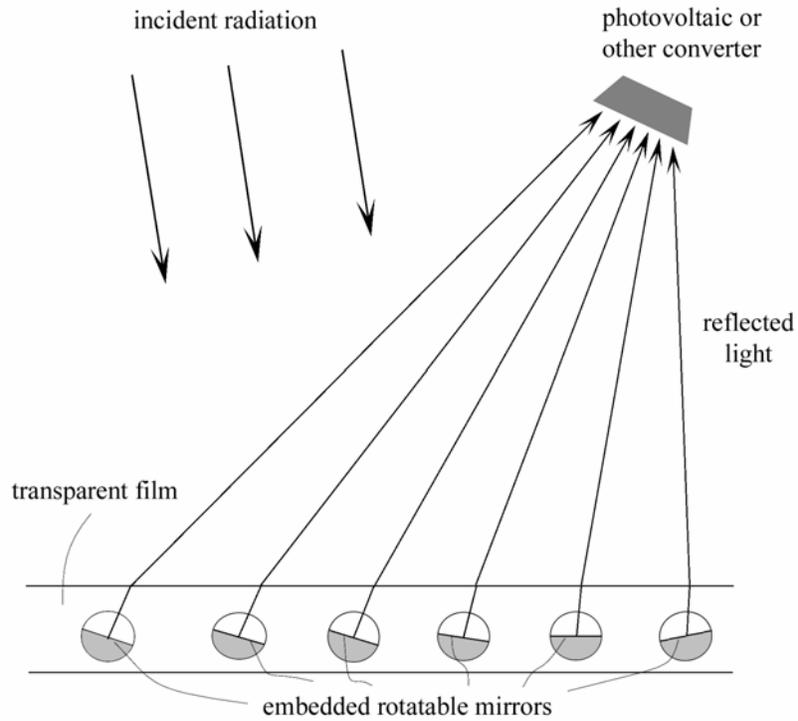

**Figure 2.** Embedded Rotatable Mirrors in a Concentrator
Showing Incident Light Reflected to a Converter.

Berkovitz, 1977) and (Sheridon et al, 1999). Resistive networks (Fig. 3) with controlled voltage nodes can be used to create an electric field that is variable in direction and magnitude.

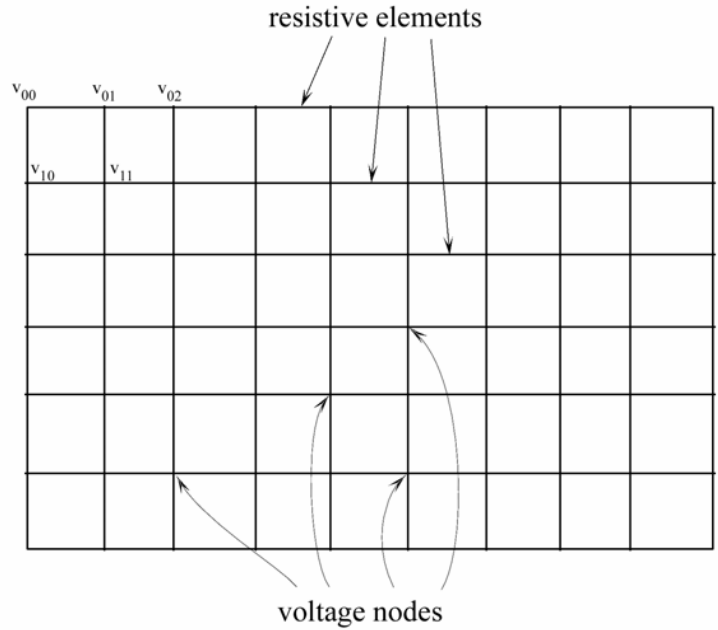

**Figure 3.** Resistive Grid to Control the Electric Field for Concentrator Ball Alignment.

The mirror array could be sandwiched between two such resistive planes or one grid and a ground plane as in Fig. 4. The plane sheets are an elastomer, mylar, silicone, or other plastic material upon which a conducting, but highly resistive material like Indium/Tin Oxide (ITO) can be sputtered to form the ground plane and/or the grid. Silicone elastomer is preferred for longevity in an ultraviolet environment. For the same reason, making the balls out of glass would be preferable. The substantial optical transparency of ITO makes it ideally suited for addressing the balls.

The orientation of the spheres is controlled by voltages $v_{ij}$ at the nodes of the grid using a small computer with analog voltage outputs. The number of mirrored spheres per grid cell would be a design variable. When there are many spheres per grid cell, the field may not be uniformly strong in the alignment direction over all the spheres. However by symmetry, this is not a problem as long as the main components of the field are in the alignment direction, and the other components cancel. The field strength only relates to how fast each sphere lines up. The cost of electronic control is rapidly decreasing. Over time and in sufficient quantities it should become a negligible fraction of the total system cost.

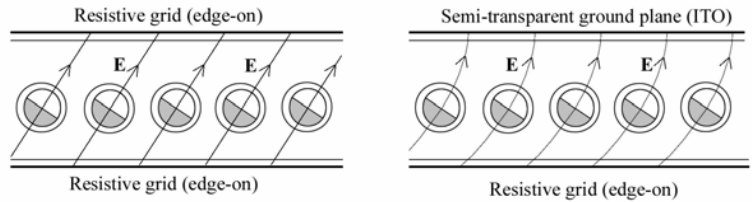

**Figure 4.** Cross-sectional View of Resistive Grid Electric Field Orientation of Mirrored Balls.

The voltage nodes would be fed to the resistive ITO grid by a backplane which would be a piece of plastic with electronic traces and possibly circuitry on it. The problem of feeding the various voltages on the grid is similar to the types of problems routinely faced in flat panel and PC board design. It is expected that optimal routing and switching techniques will be developed by electronic designers for this phase. The speeds that are required for tracking the slow motion of the sun are very modest, and so quite inexpensive forms of interconnect circuitry can be envisioned for this task.

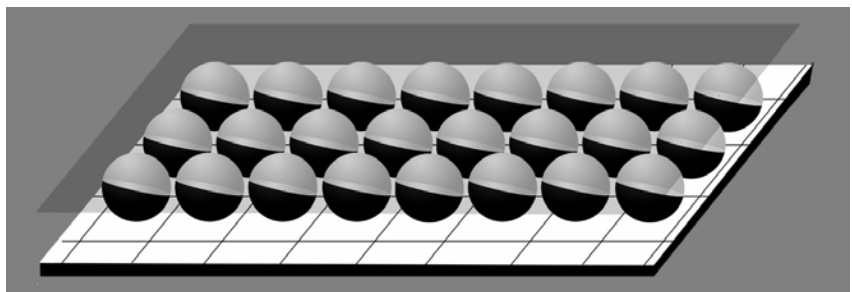

**Figure 5**. Perspective View of Mirrored Balls.

Figure 6 shows some other possible ways to implement electrically steerable micro-mirrors in rugged electronic film.

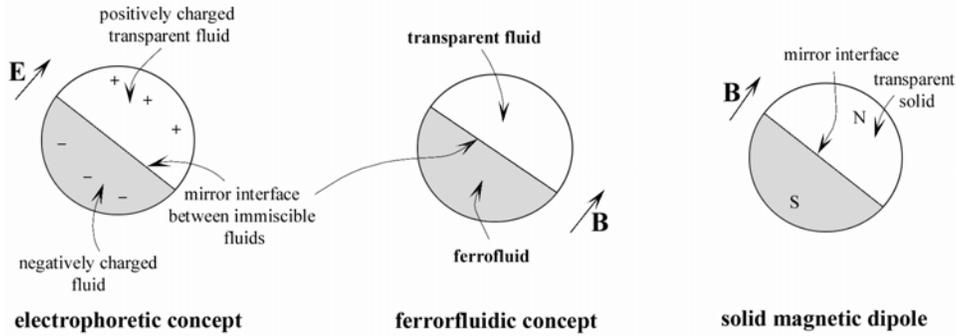

**Figure 6.** Electrophoretic, Magnetophoretic, and Magnetic Moment Mirrors in Balls.

**4. Modest concentrations and alignment accuracy**

For residential applications it would be desirable to limit the concentration level to a very safe value, ~10x. For industrial applications, this new technology can provide sufficient initial economic incentive even as low as 10x to 20x. Concentration ratios of factors of 100x to 1000x should be achievable as the technology matures. Of course at even modest concentration, a PV module requires cooling, or its performance would degrade. A demonstration that such cooling is feasible has been done at a number of facilities. For example, the Polytechnical University of Madrid has successfully operated a concentrator and PV module combination that has an enhancement of over 1000x (Swanson, 2000). Unlike the novel electronic micro-mirror concentrator, the drawback of these high enhancement concentrators is that they are bulky and require large motors and gears for tracking the sun.

Aiming the mirrors so that the reflected sunlight strikes the PV receiver is not a difficult requirement to meet at 10x. For an easy to visualize example, consider a truncated pyramid, with a 3.2 m x 3.2 m receiver, facing downward, 5 m above the center of an 11m x 11m concentrator. The aiming requirement varies from about +/- 4 degrees for the furthest mirrors at the four corners of the concentrator to about +/- 8 degrees for the mirrors nearest the receiver. Since these angles are well within the subtended angles between each mirror and the receiver, this 21% increase in concentrator area (compared to 10m x 10m) to achieve an average 10x concentration, should more than compensate for receiver shadowing, which decreases as the concentration increases. Higher concentrations have tighter tolerance requirements on mirror flatness and index of refraction matching to achieve the necessary degree of alignment. Random aiming errors can be tolerated at low concentration. If aiming errors are systematic rather than random, both for low and high concentrations, a feedback circuit detecting an imbalance of an initially balanced matrix of detectors such as thermistors on the concentrator can signal the microprocessor to compensate for the systematic error. Flux uniformity must be considered in the design of a photovoltaic system since a non-uniform flux can be detrimental to such a system. The large number of mirrors in our concentrator suggests that flux uniformity may not be a big problem. As an adjunct to a feedback circuit to enhance uniformity, random errors could be intentionally fed to the mirrors to diffuse any systematic aiming bias. Because the target area is small, it may also be cost effective to add an optical diffuser in front of the photovoltaic target. These come in many forms, and implementation would depend on the particular type of non-uniformity that is encountered.

Based upon black and white Gyricon display technology, alignment to within a few degrees is presently possible. Since the operation of our concentrator is so similar to that of commercially available Gyricon displays, we feel that a minimal experimental proof of the viability of our concept already exists -- at least for low concentrations. Color Gyricons, as well as > 10x concentrators need higher alignment accuracy. Gyricons turn off the electrical system once the balls are aligned. This may also be done after the concentrator mirrors are aligned, and they could be held in place by atmospheric pressure on the surrounding dielectric sheets, or other means. When rotation for new alignment is required, a plenum can force a small amount of the lubricating liquid in to slightly enlarge the space around each ball (Davidson and Rabinowitz, 2003). To augment programmed alignment of the mirrors, a feedback circuit can correct for both cloudy weather and any shifts in the overall concentrator orientation using the same matrix of detectors used to detect systematic mirror misalignment. In this mode, the detectors being sensitive to a decrease in solar flux can direct the microprocessor to search for a new orientation which maximizes the available signal.

**5. Blocking and other losses**

We now consider the effects of blocking (shadowing) of one mirror on another. For one layer of balls, this involves a trade-off between the packing fraction and blocking. Lowering the packing fraction decreases blocking since the mirrors are further away from each other. An optimization between these competing factors is needed to give the greatest concentration enhancement. Alternatively, more than one layer can provide near unity packing fraction, and make blocking negligible.

Aluminum is a good choice for the reflecting material because it is inexpensive and highly reflective. Other lower melting and boiling point materials that are highly reflective may be used, but are also more expensive. Only about 1000 Angstroms of Aluminum is required to get a high efficiency of reflection as is shown in Fig. 7. Figure 7 shows the unpolarized reflection from 1000 Angstroms of Al that is sandwiched between two dielectrics (e.g. a mirror sandwiched between two hemispheres), each of index of refraction 1.3. The percentage signal reflected is plotted on the vertical axis as a function of the wavelength (horizontal axis in nanometers) of the incident light. Since this is also a function of the incident angle, the angular functionality is plotted in perspective showing data points for angles from 0 degrees to 90 degrees, in the third axis. The purpose of this figure is not to give precise values for computation, rather it is to give an overview of the data. As a generalization from this figure, for most of the useful wavelengths the reflectivity is over 88%. We shall assume that transmission through the transparent materials is approximately 100%.

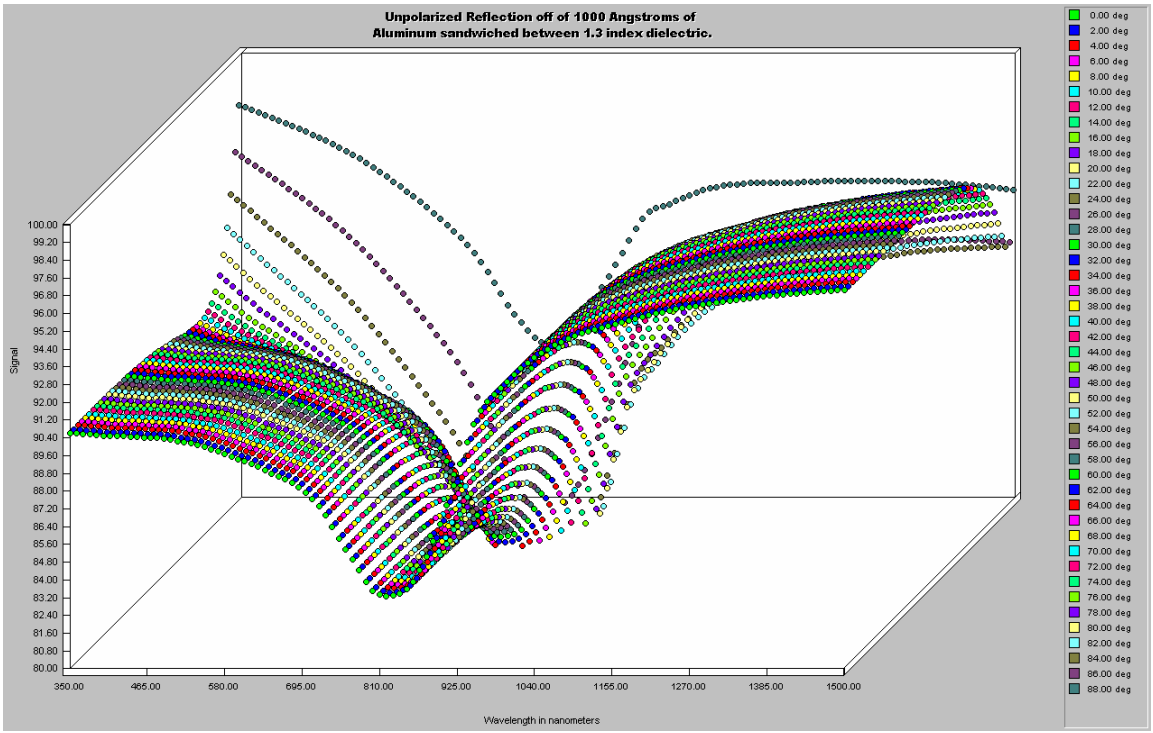

**Figure 7.** Unpolarized Reflection Dielectric over Aluminum (1000 Å). The third axis is the incident angle showing data points from 0 to 90 degrees, with 0 degrees in front and 90 degrees in back. (Elsevier Press has placed a color version of Fig. 7 on line.)

Figure 8 gives the reflection coefficients for index of refractions 1.2, 1.3, 1.4, and 1.5 as a function of angle are shown. The transmission can be calculated by subtracting these from 1.0 since there is negligible absorption at the interface. Our calculations of the transmission coefficient include transmission to the mirrors on entering the optical materials, and transmission coefficient on leaving the optical materials a second time, taking into account reflectivity. The thinness of the concentrator minimizes absorption at all wavelengths.

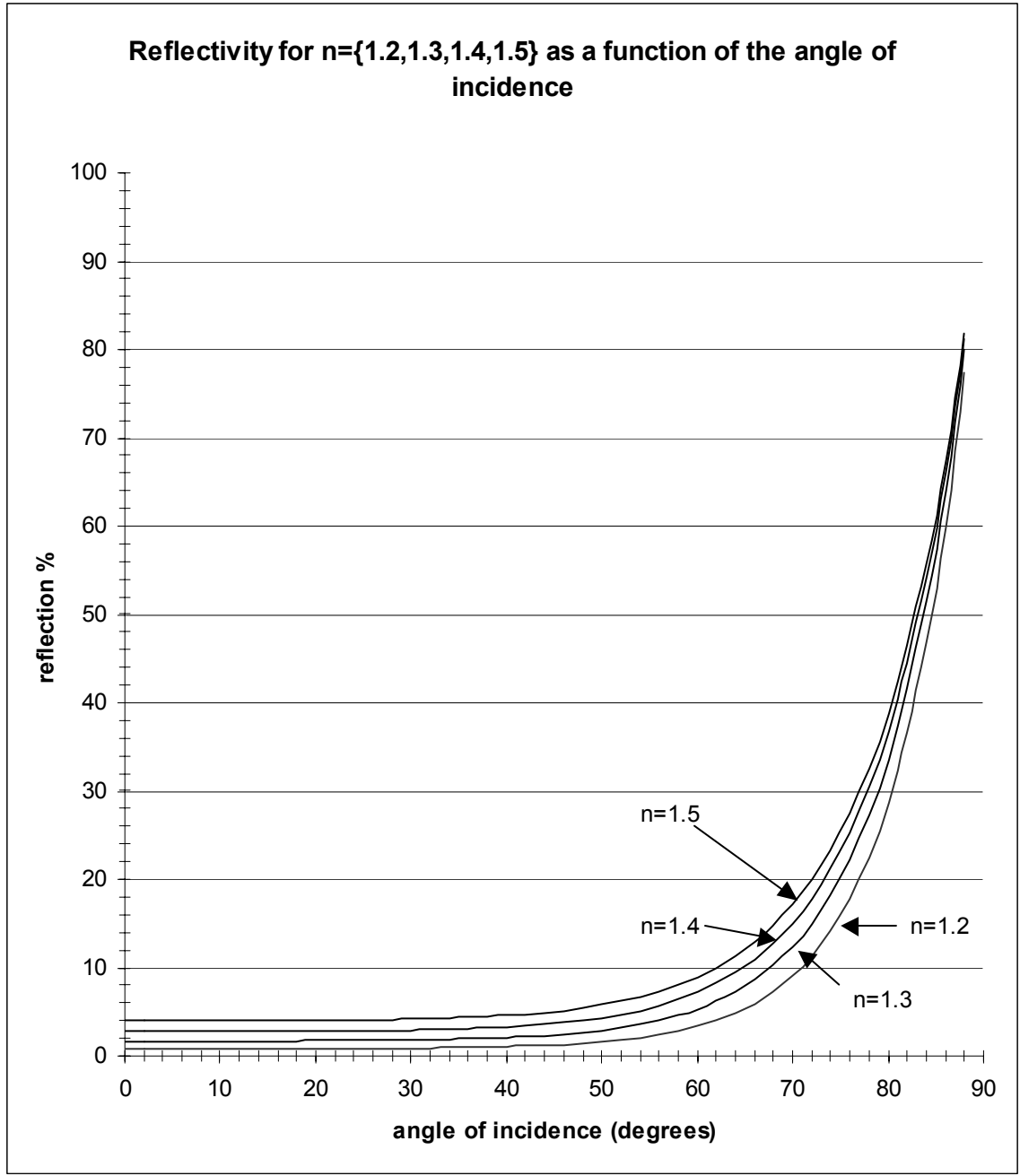

**Figure 8.** Fresnel Reflectivities of the Film for Index of Refraction n=1.2, 1.3, 1.4, and 1.5 for Unpolarized Light.

The blocking-packing density calculation we present is a **worst case approximation**. It depends on the assumption that the worst case blocking will occur when the incident ray is normal to the film and the exit ray is a grazing angle i.e. almost parallel to the plane of the film.

Blocking and shadowing losses depend on time, and may be expected to decrease as the sun moves, i.e. changes the angles. The geometry for the calculation is shown in Figure 9.

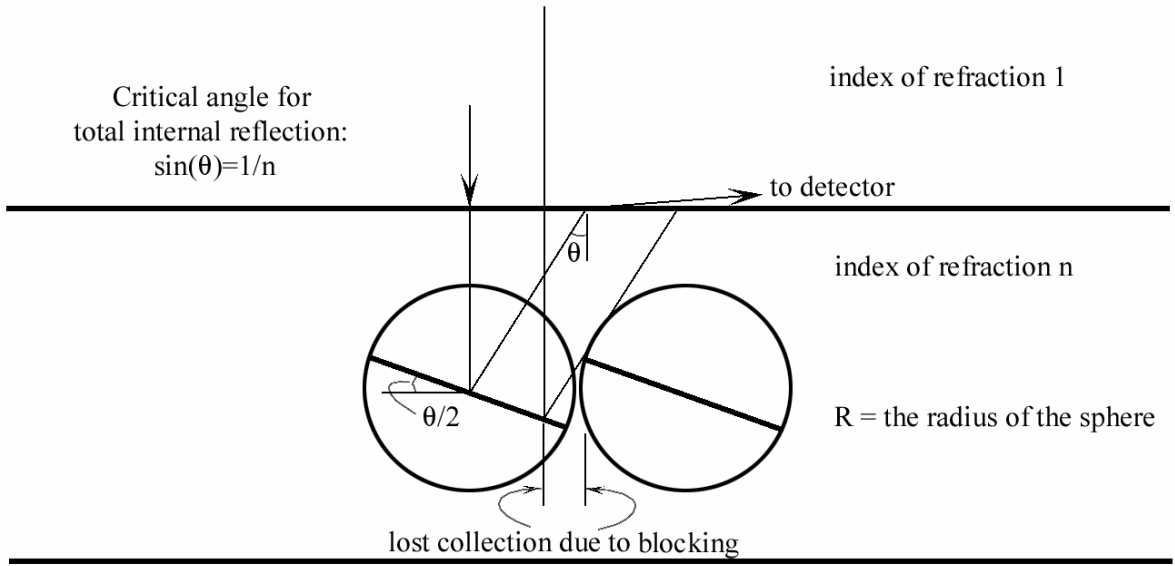

Figure 9. Efficiency Reduction due to Blocking of Light by One Mirror on Another (estimated for normal incidence and grazing angle exit ray).

The worst case combined blocking and packing density fraction occurs when:

1. The incident solar light is normal to the film. Deviation from the normal decreases the blocking.

2. The detector (receiver) is nearly in the plane of the mirror containing sheet.

We find that the packing density and blocking reduce the efficiency by a factor approximately given by $f = 1 - \sin(\theta)\tan(\theta/2)$, where $\theta < \pi/2$ since $\sin(\theta) = 1/n$, where n is the index of refraction.

| Index of refraction | Efficiency Factor f |
|---|---|
| 1.3 | 0.639 |
| 1.4 | 0.700 |
| 1.5 | 0.745 |

If the detector (receiver) is moved towards the normal to the mirror array, the efficiency factor, f, approaches 1.

In addition to the factor in Fig. 9, we include a multiplicative packing density factor of 0.91 for hexagonal packing of circles in a plane. We are now in a position to roughly estimate the overall efficiency of the concentrator. We estimate the transmission coefficients on entering and leaving the films for an overall average angle of incidence and exit of about 45 degrees.

Table 1 for the three values of n is calculated as follows. The concentrator optical efficiency including blocking effects is the product of the following factors:

A. Transmission coefficient on entering the film taking into account Reflectivity.
B. Transmission through the optical materials to the mirror.
C. Reflection off of the mirror.
D. Transmission through the optical materials a second time.
E. Transmission on leaving the film.
F. Reduction factor f due to combination of blocking and packing.
G. Packing fraction when the mirrors are parallel to the film; 0.79 for square packing, 0.91 for hexagonal packing, and 1 for > one layer.

Factors A and E: As can be seen from Fig. 8, there is only about a 3% reflection, so Factor A the transmission coefficient for light incident on the concentrator in going through the film is 1- .03 = 0.97. Similarly Factor E is 1- .03 = 0.97, since only about 3% of the light going back to the receiver is internally reflected in the concentrator.

Factors B and D: We take the incoming and outgoing transmission coefficients to be approximately 1 since the thinness of the concentrator minimizes absorption at all wavelengths.

Factor C: As can be seen from Fig. 7, for most of the useful wavelengths the reflectivity is over 88%, so Factor C is taken to be 0.9 within the accuracy of this calculation.

Factor F(f range): This reduction factor, due to a combination of blocking and packing varies as a function of angle and is given in the table in Fig. 9. For index of refraction n

= 1.3, f varies from 0.639 to 1. For n = 1.4, f varies from 0.700 to 1. For n = 1.5, f varies from 0.745 to 1.

Factor G: This is the packing fraction for one layer. We use 0.91 for hexagonal packing. For a square array, the results can be multiplied by the factor (.79/.91). If more than one layer of balls is employed, Factor G can approach 1.

| n of film | Calculation | Efficiency: min to max |
|---|---|---|
| | (A)(B =1)(C)(D =1)(E)(F f range)(G) | |
| 1.3 | (.97)(1)(.9)(1)(.97)(.639 to 1)(.91) | 0.49 to 0.77 |
| 1.4 | (.96)(1)(.9)(1)(.96)(.700 to 1)(.91) | 0.53 to 0.75 |
| 1.5 | (.95)(1)(.9)(1)(.95)(.745 to 1)(.91) | 0.55 to 0.74 |

**Table 1 (This is a worst case estimate.)**

These worst case estimates of efficiencies ranging from 49% to 77% are conservative at the lower end. One may optimistically expect 60%, but even a lower value is not that detrimental since the cost of the concentrator is the key. If it is low enough then even 49% is quite acceptable. Moreover, as a more efficient electric conversion receiver can be used economically at the focal point in a concentrator system, low concentrator efficiency can be compensated for by the higher efficiency of the receiver.

One factor that is not included but will be important is dirt on the optical surfaces which will ultimately depend on maintenance quality. The concentrator makes receiver maintenance easier, since the receiver is smaller and can face downward toward the concentrator. Where grime builds up rapidly, one may put a very thin plastic sheet over the concentrator, which is simply removed when it gets too dirty. This is much easier to do on our planar concentrator, than non-planar (e.g. parabolic) concentrators.

**6. Economic viability**

In earlier sections we have described the electronic film concentrator. Next we shall discuss its potential economic viability. We think it can make solar energy competitive even without credit for reduction in environmental pollution and health damage costs of conventional power producers. Because of our personal limited resources and the need for micro-miniaturization, making a concentrator is a slow and difficult process. It may take between 2 and 5 years to complete meaningful tests that will help to pin down efficiencies and costs. We will be glad to report our results when they are available. We emphasize that, at this time, our estimates are preliminary and have large error bars. We have made a conscientious effort to make conservative estimates to the best of our knowledge.

### 6.1 Area needed to produce 1 Megawatt peak power

The sunlight incident on the earth's atmosphere scatters in all directions. Back-scattering decreases the incident peak power density from 1.4 kW/m$^2$ above the atmosphere to about 1 kW/m$^2$ at best on the earth's surface. Though available for direct PV or thermal conversion this 1 kW/m$^2$ is further diluted by the efficiency of the system. Even 1 kW/m$^2$ is not always available for direct conversion, depending on time and weather. The diffuse component of the 1 kW/m$^2$ is not available for specular reflection from the concentrator to the receiver. So depending on atmospheric conditions, air density fluctuations and composition, only a peak power density of about 850 W/m$^2$ may be effective in a concentrator system. We use 1 kW/m$^2$ as a convenient approximate value that can easily be scaled.

Similarly, 1 MW is a convenient design parameter that can easily be scaled up for industrial applications, or down for residential applications. The area needed to meet a peak power demand of 1 Megawatt is

$$\text{Area}_{\min 1MW} = \frac{10^6 W}{F 10^3 W/m^2} = \frac{10^3 m^2}{F} = \frac{10^3 m^2}{0.1} = 10^4 m^2. \tag{1}$$

where F is the conversion efficiency. We have chosen F = 0.1, since a 10% conversion efficiency of sunlight to electricity is easily achieved.

### 6.2 Material costs

An overall film thickness of 1mm (10$^{-3}$ m) is expected for mirrored ball diameters of 200 micron (0.2 mm). Thus from Eq. (1), for a conservative efficiency of F = 0.1, the total volume of concentrators to produce 1 MW is

$$\text{Volume}_{\text{Conc 1MW}} = 10^{-3} \text{m} (10^3 \text{m}^2 / \text{F}) = \frac{1 \text{m}^3}{\text{F}} = \frac{1 \text{m}^3}{0.1} = 10 \text{m}^3. \quad (2)$$

A typical price for plastic resin in the last few years has been about $1.00/kg. The densities of most plastics are between 1 and 1.5 g/cm³ (1 to 1.5 kg/m³ ). Conservatively using the higher value, the total mass of plastic is:

$$\text{Mass}_{\text{Conc 1MW}} = 10 \text{m}^3 (1.5 \times 10^3 \text{kg} / \text{m}^3) = 1.5 \times 10^4 \text{kg}.. \quad (3)$$

Hence the total concentrator materials cost would roughly be

$$\text{Cost}_{\text{Conc 1MW}} = 1.5 \times 10^4 \text{kg} (\$1/ \text{kg}) = \$1.5 \times 10^4. \quad (4)$$

This would be about $15 per kW. With our conservative approximation of 0.1 kW/m² for the peak power, this implies a per unit area cost of

$$\text{Cost} / \text{area} = \$15 / \text{kW} (0.1 \text{kW} / \text{m}^2) = \$1.5 / \text{m}^2 \approx \$0.14 / \text{ft}^2. \quad (5)$$

Fourteen cents per square foot is an upper limit, but is only the materials cost. The balls represent a substantial fraction of the volume of the concentrator, and are preferably made of glass for longevity considerations in an ultraviolet light environment. Glass is much cheaper than plastic. However, because glass has a higher melting point and higher reactivity in the molten state, manufacturing glass balls with an internal mirror is a formidable challenge. It is expected to be much more difficult than plastic until we have mastered the process. This is why our plans include initial manufacture of plastic balls as part of the learning process. Since the main charge is induced by the physics of the environ with the balls in place, there does not appear to be a problem with getting the charge in precise alignment with each micro-mirror.

Caveats are in order. First, we are only considering peak power requirements. There is also the average power requirement which includes nighttime demand. Depending on location, the average annual power density of solar radiation on a 24 hour/day basis is typically in the range of 100–300 W/m$^2$. A number of alternatives are possible such as the production of hydrogen for fuel cells in addition to electricity; electrical and non-electrical energy storage; or augmental non-solar power must be used to meet the 24 hour/day, 365 day/year demand. Next let us consider additional costs.

**6.3 Fabrication costs of concentrator film**
It's extremely difficult to estimate the fabrication cost of the finished concentrator, but it may be expected to be within an order of magnitude of the material costs, as appears to be the case for Gyricons. A value of between 5 and 10 times material costs, or about $75/kW to $150/kW for fabrication seems reasonable with sufficient volume production. We think this is warranted for mass production as in the integrated circuit industry.

**6.4 Electronics costs**
It is reasonable that the cost of the microprocessor plus memory could be $10 per kW of produced solar power with a microprocessor per 10 m$^2$ concentrator module. The associated analog voltage control of the grid is more complicated. It might be advantageous to embed some active logic circuitry on the network grid. By keeping currents very low in the resistive tracking grid, the ohmic losses can be minimized.

**6.5 Photovoltaic conversion cost**

An excellent source for considering present PV state-of-the-art technology, and the potential for cost reductions in the next few decades is van der Zwaan and Rabl (2003). They analyze PV production cost ranges for both installation and generation, for single crystal, multi-crystalline, amorphous silicon and other thin film technologies. Table 2 shows some typical cost breakdowns of current photovoltaic (PV) systems (Dunlop et al, 2001).

| *Year 2000 Status* | |
|---|---|
| PV Module Costs | $4,000 to $8,000 per kW |
| Inverter Costs | $750 to $1,300 per kW |
| Balance of System Costs | $200 to $4,000 per kW |
| Labor Costs | $250 to $4,000 per kW |
| Total System Cost | $6,500 to $17,000 per kW[a] |
| Operating and Maintenance Costs | $.01 to $.20 per kWhr |
| [a] The low and high values do not all occur together in the same sample case and so the total system cost lies between the two extremes obtained by summing the upper and lower values. | |

**Table 2**

In 1999, the U.S. PV industry developed a 20-year roadmap which set challenging goals for increasing the volume of production and reducing costs to end-users (Dunlop et al, 2001). The roadmap calls for the following costs to be met.

| *Year 2020 targets* | |
|---|---|
| PV Module Costs | $1,000 per kW |
| Inverter Costs | $200 per kW |
| Balance of System Costs | $200 per kW |
| Labor Costs | $200 per kW |
| Total System Cost | $1,500 per kW[a] |
| Operating and Maintenance Costs | $.01 per kWhr |
| [a] The total system cost is slightly less than the sum of the targets. As these are goals, and some of these costs may actually be lower than the target, this reflects some slight optimism due to this effect. | |

**Table 3**

Some of these goals are already close to being met. For example, some manufacturers are now selling solar PV modules for only $2500/kW compared to $6000/kW in the recent past. Recent cost data for concentrator-PV module systems are currently unknown, but may hopefully follow similar cost reductions.

The advent of low cost concentrator technology can significantly impact these numbers especially in the near future. We see from Table 2 that PV module costs are presently between about 50% and 80% of the total. We see from Table 3 that the PV module costs in 2020 are expected to be 2/3 of the total cost. As calculated above, our estimated cost is

10x Concentrator Cost = Materials + Fabrication + Electronics + Voltage Control

$$= (\$15 + [\$75 \text{ to } \$150] + \$10 + \$10)/kW = \$110 \text{ to } \$185/kW. \quad (6)$$

The number of photovoltaic modules i.e. the PV area is accordingly roughly reduced in inverse proportionality to the area of the concentrator. Conservatively assuming that the cost of the photovoltaic modules used in conjunction with a electronic film 10x concentrator is about 1/5 (rather than 1/10) of the cost of ten times as many unconcentrated PV modules, or about $200/kW by the year 2020 (cf. Table 3), we see that a target cost for the PV system plus concentrator of $310 to $385 /kW is not unreasonable in the near future. One may expect the concentrator system to accelerate the decrease in PV module costs as it increases the market. In addition, the Labor costs and Balance-of-System costs will be lower since installing a rugged flexible plastic film is easier than to install fragile crystalline silicon photovoltaic panels. Assuming only a modestly conservative decrease in Balance-of-System and Labor costs:

Total System Cost = PV Module + 10x Concentrator + Balance of System + Labor

$$= (\$200 + [\$110 \text{ to } \$185] + \$150 + \$150)/kW = \$610 \text{ to } \$685/kW. \quad (7)$$

So a total installed cost of under $700 /kW is not out of the question with this system at 10x. For industrial use, the concentrator can simply be attached to the ground or on building roofs in small sections with a small PV module elevated over the centroid of each section. For a 10x concentrator, about 1/10 as many PV modules will be required, and for a 100x concentrator, about 1/100 as many PV modules will be required.

A 100x concentrator has the biggest impact in reducing costs at today's high PV prices of $2000 - $4000/kW for PV crystalline silicon modules, and $1000 - $3000/kW for PV thin film modules (van der Zwaan and Rabl, 2003). It would always be expected to bring more savings than a 10x concentrator, but for 100x the PV modules will require cooling. At concentrations > 100x, other energy conversion systems may be favored over PV.

For electric power generation, there are three primary types of generation requirement: baseload, intermediate load, and peak load. Baseload is typically the most capital intensive but has the lowest fuel cost, and peak power is the least capital intensive but has the highest fuel cost. Peak load in most countries occurs in the summer during the afternoon. Reliability must figure into the applicability of a given technology for each of the three types of generation. It is typical for a utility to target a loss of load probability (LOLP) of about 1 day in 10 years for a whole power grid system. Any generation equipment that will degrade this reliability figure must be backed up by other equipment. Solar generation is dependent on availability of sunlight, and concentrator solar systems are much more sensitive to clouds than unconcentrated ones. The easiest way to mitigate the effect of LOLP is to have some energy storage capability, and this is absolutely essential for baseload applications where power is required 24 hours a day. This adds to the cost of the solar generation system, and has traditionally been a significant obstacle for solar electric power. However, furious research and rapid improvement in reversible fuel cell technology is occurring now due to the anticipated demand in the automotive industry. Therefore, we can anticipate the availability of much lower cost storage technologies in the next 5 years that will also help solar competetiveness. Other types of storage that have been considered are: heat stored in eutectic salt reservoirs, chemical batteries, electrolytic production of hydrogen, flywheels, supeconducting electromagnetic field storage, pumping water uphill, hydrogen production, compressed air storage, etc. The overall energy yield of a given configuration will depend in a complicated way on latitude, atmospheric dust, concentrator orientation and design, diurnal variation, etc. But perhaps the most critical factor is simply the percent of days which have significant cloud cover. These will be very low efficiency days for solar concentrator systems, making such locations unsuitable for such systems. Optimizing a given concentrator configuration for a given location will depend on whether it is intended as peak power, baseload power, or intermediate load. For peak load the system must be optimized to provide the most energy in the summer months in the afternoons, but for baseload systems it must be optimized for average annual energy production. As a crude approximation, we might consider that the total radiant energy falling onto a planar array of mirrors is proporional to the cosine of the ray to the normal of the plane. The average value of the cosine from 0 to 90 degrees is 64%. This would be a very crude estimate of the ratio one would expect between the conversion efficiency for baseload verses peak power applications.

    Our cost estimates are preliminary, and it is premature to estimate an error bar for them. However, it is clear that this technology has the potential to significantly lower both the peak power and average power costs of solar photovoltaic generating systems. Our analysis was done for peak power. For average power, the costs for both PV systems with and without concentrators would be a factor of 3 to 10 times higher than the peak power costs since the average annual power density of solar radiation on a 24 hour/day basis is typically in the range of 100 - 300 W/m2. However the germane issue of providing average annual power at lowest unit cost is a site specific problem that would be difficult even for the far simpler case of peak power costs. Maximizing peak power is not sufficient for maximizing average annual power as time, geographic locale, and weather involving diurnal insolation, cosine effects, shading and blocking as a function of the sun's incidence angles, etc. need to be properly accounted for. At a given design point, the average power production during an average clear day may be expected to vary from a low of 20% to a high of 60% of the peak power depending on location, requiring up to 5 times more concentrator area for comparable power production. Thus on a superficial level, it would appear that because of it's inherently lower power density solar needs much more area than

conventional sources even when solar power can compete economically with them. Actually the playing field becomes much more level when all the multitude of non-apparent areas needed for fossil fuel sources are taken into consideration, such as mines, wells, fuel processing plants, cooling sources including rivers and the atmosphere over considerable areas and volumes, fuel storage and distribution facilities, security deployment, etc.

**7. Other system concepts and advantages**

Solar residential and/or community-based solar energy conversion is a form of distributed power that decreases vulnerability to terrorism compared to present central large power generating plants. In general solar energy enhances the security of the energy supply since it is a natural resource not confined to national borders.

Besides residential photovoltaic electricity production, the existence of a low-cost electronic film solar concentrator has a number of other applications.

1. Perhaps the most important is solar farms that can supply low cost electricity and/or hydrogen for fuel cells.
2. The high temperatures which are achievable from high concentration systems could be used in the chemical industry for heat-energy intensive processes when high aiming accuracy is achieved.
3. Solar energy could be used to augment fossil fuels in combined-cycle electricity generation plants when high aiming accuracy is achieved.
4. Artistically versatile and flexible solar lighting effects could be implemented using the focusing electronic film.

**8. Conclusions**

New technologies must accomplish at least three goals in order to have an impact on society. They must fill an important need, and must further be economically and technically viable. The micro-optics solar energy concentrator can achieve these three goals. It is less expensive than conventional solar concentrators for two reasons. First due to miniaturization, the amount of material needed for the optical system is much less. Second, because the micro-optics solar concentrator is light-weight and flexible, it can easily be attached to existing structures. This is a great economic advantage over all existing solar concentrators which require the construction of a separate structure to support them and heavy machinery to orient them to intercept and properly reflect sunlight onto a receiver.

Presently the costs of solar photovoltaic power and solar thermal power are too high to consider them seriously as contenders on the economic playing field. However, with the potential for a cost less than $700 per kW installed generating capacity, the electronic film concentrator can make solar a major energy competitor because it introduces a multitude of new design options and possibilities.

**Acknowledgements**
The authors would like to thank Laverne Rabinowitz and Joyce Davidson for their invaluable support during the course of this work; and Joe Crowley, Bernard Haisch, Joakim Lindblom, and Armand Neukermans for their helpful comments.

*Corresponding author: Armor Research, Redwood City, CA 94062-3922  USA
Mario715@earthlink.net; fax 650, 368-4466